\documentclass[12pt]{article}
\usepackage{amsmath}
\usepackage{psfrag,epsf}
\usepackage{graphicx} 
\usepackage{bmpsize}
\usepackage{enumerate}
\usepackage{natbib}
\usepackage{url} 
\usepackage{lineno} 
\usepackage{amssymb}

\newtheorem{theorem}{Theorem}[section]  

\newcommand{\blind}{0}

\begin{document}



\def\spacingset#1{\renewcommand{\baselinestretch}%
{#1}\small\normalsize} \spacingset{1}


\if0\blind
{
  \title{\bf A Constructive Spatio-Temporal Approach to Modeling
    Spatial Covariance}
  \author{Ephraim M. Hanks\thanks{
    The author gratefully acknowledges conversations with John Fricks,
    whose suggestions were instrumental in the analysis of Section 3.1}\hspace{.2cm}\\
    Department of Statistics, The Pennsylvania State University\\
}
  \maketitle
} \fi

\if1\blind
{
  \bigskip
  \bigskip
  \bigskip
  \begin{center}
    {\LARGE\bf A Constructive Spatio-Temporal Approach to Modeling
    Spatial Covariance}
\end{center}
  \medskip
} \fi

\bigskip
\begin{abstract}
I present an approach for modeling areal spatial covariance by
considering the stationary distribution of a spatio-temporal Markov
random walk.  This stationary distribution corresponds to an intrinsic
simultaneous autoregressive (SAR) model for spatial correlation, and
provides a principled approach to specifying areal spatial models when
a spatio-temporal generating process can be assumed.  I apply the
approach to a study of spatial genetic variation of trout in a stream
network in Connecticut, USA, and a study of crime rates in
neighborhoods of Columbus, OH, USA.
\end{abstract}

\noindent%
\vfill

\newpage

\spacingset{1.45} 

\section{Introduction}
\label{sec:intro}

Almost all spatial data can be viewed as arising from a
spatio-temporal generating process.  For example, a spatial survey of infectious
disease prevalence is a snapshot of a dynamic epidemic process
occuring in space and time.  Similarly, spatial genetic data are the result of
spatio-temporal dispersal, mating, and survival processes at the
population level.  When these spatial processes are observed at multiple
successive time points, the known science behind the spatio-temporal
process is often used to motivate a spatio-temporal statistical model
\citep[e.g.,][]{Wikle2010,CressieWikleBook}.  

In contrast, consider the case of ``spatial'' data, where only one temporal realization of the
spatio-temporal process is observed.  In this case, spatial autocorrelation is often modeled
by including a spatial random effect \citep[e.g.,][]{Diggle2007} in
the fitted statistical model.  The
prior distribution for this spatial random effect is almost always
modeled semiparametrically using a Gaussian process model
with covariance
function chosen based on the support of the data, irrespective of the 
spatio-temporal generating process.  For example, when the spatial
data are point-referenced, the Matern class of covariance functions
\citep[e.g.,][]{Cressie1993} are often used, while if the spatial data
have areal or lattice support, then either conditional autoregressive
\citep[CAR; e.g., ][]{Besag1974, Besag1995, RueText} or simultaneous
autoregressive \citep[SAR: e.g., ][]{Wall2004, CressieWikleBook}  
models are common.  In either case, the choice of prior distribution 
for the spatial random effect is almost always made based solely on
the support of the data, without consideration of an underlying
generating process.

Spatial data are poor in
information relative to spatio-temporal data; however, we are
increasingly able to collect large amounts of spatial data.  The
increased information present in large spatially-correlated data
provides an opportunity for more realistic modeling of spatial
covariance than has been
possible in the past.  Additionally, recent recognition of the
potential for spatial confounding
\citep{HodgesReich2010,Paciorek2010,Hughes2013,HanksRSR} highlights
the need to choose a spatial model with care, as the structure of a
spatially-correlated random effect can influence inference on fixed
effects.  


I propose a general constructive approach to modeling spatial
correlation based on considering the stationary distribution of a
spatio-temporal generating process.  This 
spatio-temporal generating process can either be specified based on
scientific knowledge, or can be thought of simply as a device to
construct a spatial correlation with desired properties, such as
anisotropy and nonstationarity.  In Section 2, I describe the
proposed general approach, and link it to current spatial models for 
 continuous (geostatistical) random fields.  In
Section 3, I focus on areal spatial models and show that the
stationary distribution for a
spatio-temporal random walk model results in a spatial SAR model,
which provides a principled approach for choosing areal neighbors and
SAR weights when spatial data can be seen as arising from a
spatio-temporal random walk.  In Section 4, I use this development to
model spatial genetic data based on a spatio-temporal random
walk generating process.  I apply this model to genetic data
collected from trout in the Jefferson-Hill Spruce Brook in
Connecticut, USA.  In Section 5 I present a second example by modeling
crime rates for areal neighborhoods in Columbus, Ohio, USA.  This
second example illustrates how a spatio-temporal generating process
can be used to jointly model fixed and random spatial effects.  In
Section 6 I close with discussion of the proposed approach.




\section{A Constructive Spatio-Temporal Approach to Modeling Spatial Covariance}

The proposed approach is as follows.
\begin{enumerate}
\item Define a deterministic spatio-temporal generating model for the
  spatio-temporal process $\mathbf{y}(s,t)$, where $s$ indexes space
  and $t$ indexes time
\begin{equation}
\frac{\partial}{\partial t} \mathbf{y}(s,t)=\mathcal{F}\left(
\mathbf{y}(s,t)\right).
\end{equation}
For example, $\mathcal{F}$ could be a differential operator (e.g.,
$\frac{\partial^2}{\partial s^2}$) in which case (1) is a partial
differential equation (PDE). 
\item Drive the spatio-temporal process defined by (1) with time-homogeneous spatial
  noise $\mathbf{W}(s)$
\begin{equation}
\frac{\partial}{\partial t} \mathbf{y}(s,t)=\mathcal{F}\left(
\mathbf{y}(s,t)\right)+\mathbf{W}(s)\quad,\quad \mathbf{W}(s) \sim
N(\cdot,\cdot).
\end{equation}
The process (2) is now a random (stochastic) process in contrast to
(1), which is deterministic.
\item The stationary distribution
  $\boldsymbol\pi(s)=\lim_{t\rightarrow \infty} \mathbf{y}(s,t)$
  of (2) provides a spatial model capturing the 
  dynamics of the spatio-temporal process.
\begin{equation}
\frac{\partial}{\partial t} \mathbf{y}(s,t)=0 \quad \Rightarrow\quad
\mathcal{F}\left(
\boldsymbol\pi(s)\right) = -\mathbf{W}(s)
\end{equation}
\end{enumerate}

Solving (3) for the stationary distribution $\boldsymbol\pi(s)$ can be
done analytically in some cases, but in many others a numerical
approximation will be required.

\subsection{Spatio-Temporal Generating Models for Continuous-Space Spatial Models}

I first consider this approach in the context of continuous-space processes, and restrict attention
to spatial processes in $\mathcal{R}^2$, 
with the two dimensions $\mathbf{s}=(x_1,x_2)$.  The
generalization to higher (or lower) spatial dimensions is
straightforward \citep{Lindgren2011}.  The most common spatial
covariance function used in continuous space is the Matern class, with
covariance function given by
\[\text{cov}(\mathbf{s}_i,\mathbf{s}_j)=\sigma^2\frac{1}{\Gamma(\nu)2^{\nu-1}}
\left(\sqrt{2\nu}\frac{d_{ij}}{\phi}\right)^\nu
K_\nu\left(\sqrt{2\nu}\frac{d_{ij}}{\phi}\right)
\]
where $d_{ij}=\sqrt{(x_{i1}-x_{j1})^2+(x_{i2}-x_{j2})^2}$ is the Euclidean distance between the spatial locations
of the $i$-th and $j$-th observations, 
$\sigma^2$ is the partial sill parameter, $\nu$ is the Matern
smoothness parameter, $\phi$ is a range parameter, and $K_\nu(\cdot)$
is the modified Bessel function of the second kind
\citep[e.g.,][]{Cressie1993}.  

As a special case of the
constructive spatio-temporal approach proposed in the previous section, consider the random partial
differential equation  
\begin{equation}
\frac{\partial}{\partial t}
\mathbf{y}(x_1,x_2,t)=(\Delta-\kappa^2)^{\alpha/2}\mathbf{y}(x_1,x_2,t)
+\mathbf{W}(x_1,x_2),
\end{equation}
where $\Delta=\frac{\partial^2}{\partial x_1^2}+\frac{\partial^2}{\partial x_2^2}$ is the Laplacian and $\mathbf{W}(x_1,x_2)$ is
time-homogeneous spatial Gaussian white noise.  Note that while equation (4)
has been termed a stochastic partial differential equation
\citep[SPDE; ][]{Lindgren2011}, I follow \cite{KloedenPlaten} and
reserve SPDE to refer to a differential equation model driven by noise
which varies over time (e.g., $\mathbf{W}(x_1,x_2,t)$ could be a spatial
Wiener process), while a random partial differential equation (RPDE)
is a differential equation driven by time-homogeneous noise, as in (2)
and (4).

The stationary distribution of (4) satisfies the RPDE 
\[(\kappa^2-\Delta)^{\alpha/2}\boldsymbol\pi(x_1,x_2)=\mathbf{W}(x_1,x_2),\]
whose solution is a random field of the Matern class
\citep{Whittle1954,Lindgren2011}.  As a concrete example, consider (4)
when $\kappa^2=0$ and $\alpha=2$
\begin{equation}
\frac{\partial}{\partial t}
\mathbf{y}(x_1,x_2,t)=\left(\frac{\partial^2}{\partial x_1^2}+\frac{\partial^2}{\partial x_2^2}\right)\mathbf{y}(x_1,x_2,t)+\mathbf{W}(x_1,x_2).
\end{equation}
The spatio-temporal generating
process (5) is a two-dimensional diffusion with time-homogeneous spatial sources
and sinks defined by $\mathbf{W}(s)$.  The corresponding stationary
spatial distribution is an intrinsic Matern random field with
smoothness parameter $\nu=2$ \citep[See ][for details]{Lindgren2011}.

The novelty introduced in this section is the temporal
aspect of the RPDE approach, which is not necessary to use spatial
models motivated by RPDEs.  However, the interpretation of spatial
models as stationary distributions of spatio-temporal RPDEs opens up
the possibility of scientific modeling of spatial random effects when
a spatio-temporal generating process, such as diffusion (5), can be
assumed.  Having shown how the Matern class of spatial models can be
seen as the stationary distributions of the continuous space
spatio-temporal RPDE (4), 
I now consider spatio-temporal models with discrete (areal) spatial support. 


\section{Discrete Space Random Walk Models for Spatial Covariance}
\label{sec:disc}

\cite{Lindgren2011} consider discrete (areal) spatial models in the
context of numerically approximating the solution to the RPDE (4)
using a finite element basis set.  Approximating a continuous
spatial random field with a finite element approximation leads to
increases in computational efficiency, as the finite element basis
solution results in a Gaussian Markov random field (GMRF) with sparse
precision matrix.  

Instead of continuous spatial effects, consider modeling
areal spatial processes directly.  That is, consider modeling a
spatial random effect $\mathbf{y}=[y_1,y_2,\ldots,y_n]'$ on $n$
spatial locations which constitute the full spatial support of the
random effect.  While both CAR and SAR models have been used
extensively to model spatial autocorrelation in areal spatial models,
there is little in the way of guidance on when to use one or the
other, and little guidance on how to define the neighborhood structure
that defines the CAR and SAR models \citep[e.g.,][]{Wall2004,Assuncao2009}.

By considering a Markov random walk on a discrete spatial support, I
will first derive a population-level diffusion RPDE based on a large-population
approximation to the spatial movements of many individuals.  I will
then show that the stationary distribution to this RPDE is a SAR
model.  This result will then be used in Section 5 and Section 6 to propose  
spatial covariance models based on population-level spatial random
walk or diffusion processes.

\subsection{Population-level Markov random walks}

Let $\mathbf{G}=(\mathbf{V},\mathbf{E})$ be a graph with vertices
$\mathbf{V}=\{V_1,V_2,\ldots,V_M\}$ and directed edges
$\mathbf{E}=\{\alpha_{ij},i=1,2,\ldots,M;j=1,2,\ldots,M\}$.  In
particular, consider the case where $\alpha_{ij}$ is the
exponential rate at which a random walker in node $i$ transitions to
node $j$.  As in a standard continuous-time Markov chain (CTMC) model
for a random walk, the time $T_i$ spent by a random walker in node $i$
before transitioning to any other node is exponentially-distributed
with rate $\alpha_i=\sum_{k=1}^n \alpha_{ik}$.  

Consider population-level processes on the graph $\mathbf{G}$ in which there
are $N$ members of the population, all behaving as a random walk.
If there are $n_i(t)$ individuals at node $i$ and time $t$, then
the rate at which individuals move from node $i$ to node $j$ is given
by $n_i\alpha_{ij}$.  Following \cite{Kurtz1978} and
\cite{Baxendale2011}, the normalized population
process $\mathbf{z}(t)=[z_1(t) \ z_2(t)\ \ldots z_M]$ can be defined
as 
$z_i(t)=n_i(t)/N$.  

\begin{figure}[tb]
\begin{center}
\includegraphics[width=5.1in,natwidth=337,natheight=231]{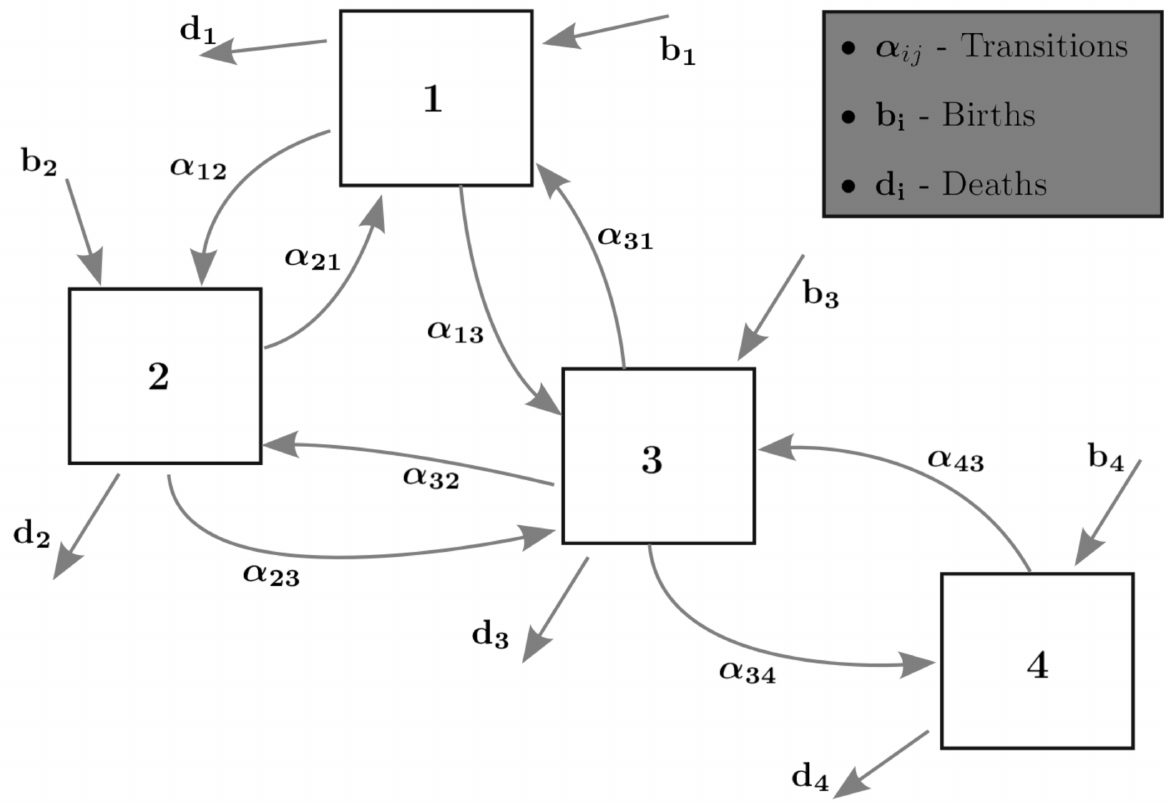}
\end{center}
\caption{Continuous-time Markov random walk model example.
  $\alpha_{ij}$ is the transition rate from node $i$ to node $j$ and
  may be zero, indicating direct migration is impossible without
  traversing other nodes.  $b_i$ is the rate at which individuals are
  introduced into the system at node $i$, and $d_i$ is the rate at
  which individuals in node $i$ are removed from the system.}
\end{figure}

In an open population model, individuals may enter (birth) or leave
(death) the system continuously in time at any node (Figure 1).  It is common to model the birth and
death rates at node $i$ as being density dependent, with birth rate of $n_i b$ and
death rate of $n_i d$, for constant rates $b$ and $d$ shared across
space.  Instead, I will allow the birth and death rates to vary
spatially, as this will provide a convenient mechanism for accounting
for unmodeled spatial variation.  To this end, consider birth and
death rates that scale with the total population size ($N$).  Let $N
b_i$ be the rate at which 
individuals are introduced into node $i$ and let $N d_i$ be the rate
at which individuals in node $i$ are removed from the system.   

To write a spatio-temporal model for the normalized population process
$\mathbf{z}(t)$, it will be helpful to write each of the potential
jumps (movement between nodes, births, and deaths) possible in this
discrete system.  If an individual is introduced at node $i$, then the
population at $i$ increases by 1.  Notationally, represent this transition in
the population process $\mathbf{n}$ as $\mathbf{n} \rightarrow
\mathbf{n}+\mathbf{e}_i$, where $\mathbf{e}_i$ is the canonical vector with $M$ componants,
all of which are zero except for the $i$-th element, which is equal to
1.  The jump in this birth transition is given by $\mathbf{e}_i$.  Similarly, a death (removal)
at node $i$ decreases the population at node $i$ by 1 and is given by
the jump $-\mathbf{e}_i$.  Spatial movement (transitions) from node $i$
to node $j$, which occur with rate $n_i\alpha_{ij}$, have jumps given
by $\mathbf{e}_i-\mathbf{e}_j$.  The possible transitions with their
rates are given in Table 1.

\begin{table}[ht]
\caption{Transitions and Poisson rates in the continuous-time Markov
  population process.  \label{tab:tabone}}
\begin{center}
\begin{tabular}{llll}
Description & Transition & Jump & Rate \\\hline
Birth at node $i$ & $\mathbf{n} \rightarrow
\mathbf{n}+\mathbf{e}_i$ & $\mathbf{e}_i$ & $Nb_i$  \\
Death at node $i$ & $\mathbf{n} \rightarrow
\mathbf{n}-\mathbf{e}_i$ & $-\mathbf{e}_i$ & $Nd_i$  \\
Move from node $i$ to node $j$ & $\mathbf{n} \rightarrow
\mathbf{n}+\mathbf{e}_j-\mathbf{e}_i$ & $\mathbf{e}_j-\mathbf{e}_i$ & $n_i\alpha_{ij}$  \\
\end{tabular}
\end{center}
\end{table}

Given an initial unnormalized population state $\mathbf{n}(0)$ at time
zero, the transient distribution $\mathbf{n}(t)$ is given by
\citep[e.g.,][]{Baxendale2011}
\[\mathbf{n}(t) = \mathbf{n}(0)+ \sum_{ij \neq 0} (\mathbf{e}_{j}-\mathbf{e}_i)
P_{ij}\left[ \int_0^t n_i(s)\alpha_{ij}\text{d}s\right] 
+ \sum_i \left( \mathbf{e}_{i} P_{0i}\left[ \int_0^t Nb_i\text{d}s\right] -\mathbf{e}_i 
P_{i0}\left[ \int_0^t Nd_i\text{d}s\right]\right)
\]
where \[P_{ij}(a) \sim Pois(a),\quad i=0,1,\ldots,M;\ 
j=0,1,\ldots,M; \ i\neq j.\]
The transient distribution for the normalized density
$\mathbf{z}=\mathbf{n}/N$ is given by
\begin{equation}
\mathbf{z}(t) = \mathbf{z}(0)+ \sum_{ij \neq 0}
(\mathbf{e}_{j}-\mathbf{e}_i) \frac{1}{N}
P_{ij}\left[ \int_0^t n_i(s)\alpha_{ij}\text{d}s\right] 
+ \sum_i \mathbf{e}_i   \left(\frac{1}{N}P_{0i}\left[ Nb_i t\right] - \frac{1}{N}P_{i0}\left[ Nd_i t\right]\right).
\end{equation}
Taking the large population limit as $N\rightarrow\infty$  \citep{Kurtz1978,Baxendale2011} gives the
integral equation for the normalized density
\begin{equation}
\mathbf{z}(t) = \mathbf{z}(0)+ \sum_{i \neq j } (\mathbf{e}_{j}-\mathbf{e}_i)
\int_0^t z_i(s)\alpha_{ij}\text{d}s + \sum_i
\mathbf{e}_{i}  (b_i-d_i) t.
\end{equation}
Details of this calculation are given in Appendix A.

The differential equation associated with (7) is 
\[
\frac{\partial \mathbf{z}(t)}{\partial t} = \sum_{i \neq j } \alpha_{ij}(\mathbf{e}_{j}-\mathbf{e}_i)
z_i(t) + \sum_i \mathbf{e}_{i}  (b_i-d_i)
\]
which has vectorized form
\begin{equation}
\frac{\partial \mathbf{z}(t)}{\partial t} = -\mathbf{Q}'\mathbf{z}(t) +(\mathbf{b} -\mathbf{d}) 
\end{equation}
where $\mathbf{b}=[b_1\ b_2\ \ldots \ b_M]'$, $\mathbf{d}=[d_1\ d_2\
\ldots \ d_M]'$, and $\mathbf{Q}$ is the infinitessimal generator of
the CTMC or the Laplacian matrix of the graph
\begin{equation}
\mathbf{Q}=\left(\begin{array}{ccccc}
\sum_k \alpha_{1k} & -\alpha_{12} & -\alpha_{13} & \cdots &
-\alpha_{1m} \\
-\alpha_{21} & \sum_k \alpha_{2k} &  -\alpha_{23} & \cdots &
-\alpha_{2m} \\
-\alpha_{31} & -\alpha_{32} &\sum_k \alpha_{3k}  & \cdots &
-\alpha_{3m} \\
\vdots & &   &\ddots & \vdots \\
-\alpha_{m1} &  -\alpha_{m2} & -\alpha_{m3} & \cdots &
 \sum_k \alpha_{mk}
\end{array}\right).
\end{equation}

Equation (8) specifies a graph diffusion process where
$\mathbf{b}-\mathbf{d}$ is a vector of net inputs and outputs to the
system and $-\mathbf{Q}'$ is a matrix describing proportional rates of
transfer between spatial locations.

\subsection{Spatial Models From Random Walks}

To specify a spatial model motivated by the differential equation (8),
consider modeling the spatial birth and death rates as spatial white
noise
\[\boldsymbol\gamma=\mathbf{b}-\mathbf{d} \sim
N(\mathbf{0},\sigma^2\mathbf{I})\] 
subject to the constraint that $\mathbf{1}'\boldsymbol\gamma=0$.  This
sum-to-zero constraint on $\boldsymbol\gamma$ is necessary to ensure
the existence of a stationary distribution $\boldsymbol\pi$ for (8).
The spatio-temporal differential equation (8) can then be written as
the RPDE
\begin{equation}
\frac{\partial}{\partial t} \mathbf{z}(t) =
-\mathbf{Q}'\mathbf{z}(t)+\boldsymbol\gamma,\quad
\boldsymbol\gamma\sim N(\mathbf{0},\sigma^2\mathbf{I}).
\end{equation}
The stationary distribution $\boldsymbol\pi$ for the normalized
population process $\mathbf{z}$ satisfies the balance equation that
$\frac{\partial}{\partial t} \mathbf{z}(t) =\mathbf{0}$, which implies
that 
\[\mathbf{Q'}\boldsymbol\pi = \boldsymbol\gamma,\quad
\boldsymbol\gamma\sim N(\mathbf{0},\sigma^2\mathbf{I})\]
and thus the stationary distribution for (10) is given by
\begin{equation}
\boldsymbol\pi \sim N(\mathbf{0}, (\mathbf{QQ}')^-), \text{ with
}\mathbf{1}'\boldsymbol\pi=0.
\end{equation}

This stationary distribution is a random field on the discrete spatial
support of the population process $\mathbf{z}(t)$ with spatial
covariance defined by the spatio-temporal CTMC random walk with
infinitessimal generator $\mathbf{Q}$ (9).

\subsubsection{Links to Intrinsic Simultaneous Autoregressive Random Fields}

The random field in (11) corresponds to
an intrinsic simulataneous autoregressive (SAR) model for spatial
correlation.  This correspondence provides an intuitive approach for
specifying the SAR neighborhood structure in situations where some
information is known about the spatio-temporal dynamics of the system
being modeled. 

The standard SAR
  model can be written \citep[see e.g., Section 4.2.7
  of][]{CressieWikleBook} as 
\[\mathbf{y}\sim
N(\mathbf{0},(\mathbf{I}-\mathbf{B})^{-1}\boldsymbol\Lambda(\mathbf{I}-\mathbf{B}')^{-1}
)\]
where $\mathbf{B}$ has zeroes on the diagonal and $\boldsymbol\Lambda$
is a diagonal matrix with $i$-th diagonal $\Lambda_{ii}$.  Then setting 
\[B_{ij}=\frac{\alpha_{ji}}{\sum_k \alpha_{ik}} \text{ and }
\Lambda_{ii}=\frac{1}{\left(\sum_k \alpha_{ik}\right)^2}\]
 expresses (6) as an
intrinsic SAR model.  As in standard SAR models, the matrix $\mathbf{Q}$ from (4) does not have
  to be symmetric, but rather can incorporate models for asymmetric
  random walks.  Additionally, if $\mathbf{Q}$ is sparse (many of the
  $\{\alpha_{ij}\}$ are zero), then sparse matrix methods
  \citep[e.g.,][]{RueText} can be
  employed to sample from and evaluate the density in (6).  

The SAR models (and related CAR models)
have been viewed as unintuitive \citep{Wall2004}.  
The spatio-temporal random walk motivation
for the spatial model in (11) provides a principled framework for incorporating
knowledge about the spatio-temporal spread of a system into a model
for spatial autocorrelation.  

The random field $\boldsymbol\pi$ in (11) is an intrinsic random
  field, in that only linear combinations are proper \citep{Besag1995}.  An
  alternative formulation is that the density
  for $\boldsymbol\pi$ is proper under the constraint that
  $\boldsymbol\pi$ sums to zero over the spatial domain.  Intrinsic
  random fields are often used as prior distributions, where the
  posterior distribution is proper.  
  For example, consider modeling a Gaussian response
  $\mathbf{y}$ as 
\[\mathbf{y} = \mu\mathbf{1} + \boldsymbol\pi +
\boldsymbol\epsilon,\quad \boldsymbol\epsilon \sim N(\mathbf{0},\tau^2\mathbf{I})\]
where $\boldsymbol\pi \sim N(\mathbf{0}, (\mathbf{QQ}')^-), \text{ with
}\mathbf{1}'\boldsymbol\pi=0.$
Under this formulation, $\boldsymbol\pi$ is constrained to sum to
zero, but $\mu\mathbf{1}+\boldsymbol\pi$ is not.  This
formulation can be seen as a form of restricted spatial regression
\citep{Hughes2013,HanksRSR} where the spatial random effect
$\boldsymbol\pi$ is constrained to be orthogonal to the intercept
$\mu\mathbf{1}$.

\subsubsection{Identifiability}

The likelihood of (11) 
\[f(\boldsymbol\pi|\mathbf{Q}) \propto |\mathbf{QQ}'|^{-1/2}\exp\left\{-\frac{1}{2}\boldsymbol\pi'\mathbf{QQ}'\boldsymbol\pi\right\}\]
is a function of $\mathbf{QQ}'$, rather than purely a function of the
infinitessimal generator $\mathbf{Q}$.  Thus, if there are two
generator matrices $\mathbf{Q}$ and $\mathbf{W}$ such that 
$\mathbf{QQ}'=\mathbf{WW}'$, then
$\mathbf{Q}$ is not identifiable.  However, the special structure
required for a generator matrix of a CTMC allows us to prove that
$\mathbf{Q}$ is identifiable in all but pathological situations.


\begin{theorem}
If 
$\mathbf{Q}$ and $\mathbf{W}$ are both generator matrices (9) for
irreducible $M$-state CTMCs, and
 at least one row of $\mathbf{Q}$ has more than one nonzero off-diagonal
  entry, 
then $\mathbf{QQ}'$=$\mathbf{WW}'$ if and only if
$\mathbf{Q}=\mathbf{W}$.
\end{theorem}

The proof is given in Appendix B.
The significance of this result is that the only forms for
$\mathbf{Q}$ that are unidentifiable come when the embedded chain of
the irreducible CTMC governed by $\mathbf{Q}$ is deterministic and topologically
the graph given by $\mathbf{Q}$ is a loop, with flow only possible in
one direction (either clockwise or counter-clockwise).  In all other graph
topologies, identifiability is guaranteed.


\section{Example 1: Random walk models for spatial genetic variation on
  stream networks}

I now present two examples of spatial analyses using the assumption
of a spatio-temporal random walk generating process leading to a spatial random
effect.  The first example comes from landscape ecology, where a
common goal is to understand how the landscape 
influences spatial connectivity or correlation.  Random walk models are
among the most common models for gene flow, both in theory and in
practice.  \cite{McRae2006} showed that under a random walk
model for migration, a common formulation of genetic dissimilarity
(the linearized fixation index) was proportional to the circuit resistance
distance \citep{Klein1993} between the nodes in question.  Under the
formulation of \cite{McRae2006}, the spatial domain is envisioned as a
graph of spatial nodes with symmetric edge weights $\alpha_{ij}$
proportional to the (symmetric) rate 
of random walkers between nodes.  The
resistance distance is the effective resistance in an electric circuit
where the nodes are connected by resistors with resistance
$1/\alpha_{ij}$ equal to 
the inverse of the migration rate.  This approach to studying gene
flow is known as the isolation by resistance approach, and is
often used to explore the relationship between landscape features and
gene flow.  

While most studies addressing isolation by resistance
choose between a finite number of pre-specified edge weights
(resistances), \cite{HanksHootenJASA} modeled the observed pairwise genetic
distance matrix using the generalized Wishart
distribution of \cite{McCullagh2009} with symmetric precision matrix
$\mathbf{Q}$ (9) and made inference on the edge weights $\alpha_{ij}$
as a function of landscape covariates.  Instead of using the RPDE stationary distribution approach that
  gives rise to (11), \cite{HanksHootenJASA} considered a variogram
  argument, as follows.  Using links between symmetric random walks
  and electric circuits \citep{Doyle1984}, \cite{McRae2006} showed
  that under a random walk model for migration, a common formulation
  of genetic dissimilarity (the linearized 
  fixation index) was proportional to the resistance distance
  \citep{Klein1993}.  \cite{HanksHootenJASA} showed that the
  resistance distance was exactly the variogram (expected squared
  difference) of an intrinsic Gaussian spatial random field with precision
  matrix $\mathbf{Q}$.  While this provides an interesting link
  between random walks and variograms, our goal in this analysis is to
  directly motivate a spatial model by the stationary distribution of
  a spatio-temporal model, something not explicitly considered by \cite{HanksHootenJASA}.

The isolation by resistance approach assumes symmetric edge weights
(and thus symmetric migration rates), though often it would be more
realistic to assume asymmetric migration rates reflecting source and
sink dynamics.  As an example, consider the system studied by
\cite{Kanno2011}, consisting of trout in the Jefferson-Hill Spruce
Brook in Connecticut, USA.  470 trout were captured at 173 spatial
locations along the brook (Figure 2) and genotyped, with microsatellite allele
data obtained at 15 loci.  An isolation by resistance approach would
require symmetric migration rates between upstream and downstream
locations, but a more realistic model (which I will propose) would consider asymmetric
migration rates reflecting the potentially increased difficulty in
moving upstream relative to moving downstream.  

\begin{figure}[ht]
\begin{center}
\includegraphics[width=4in,natwidth=720,natheight=576]{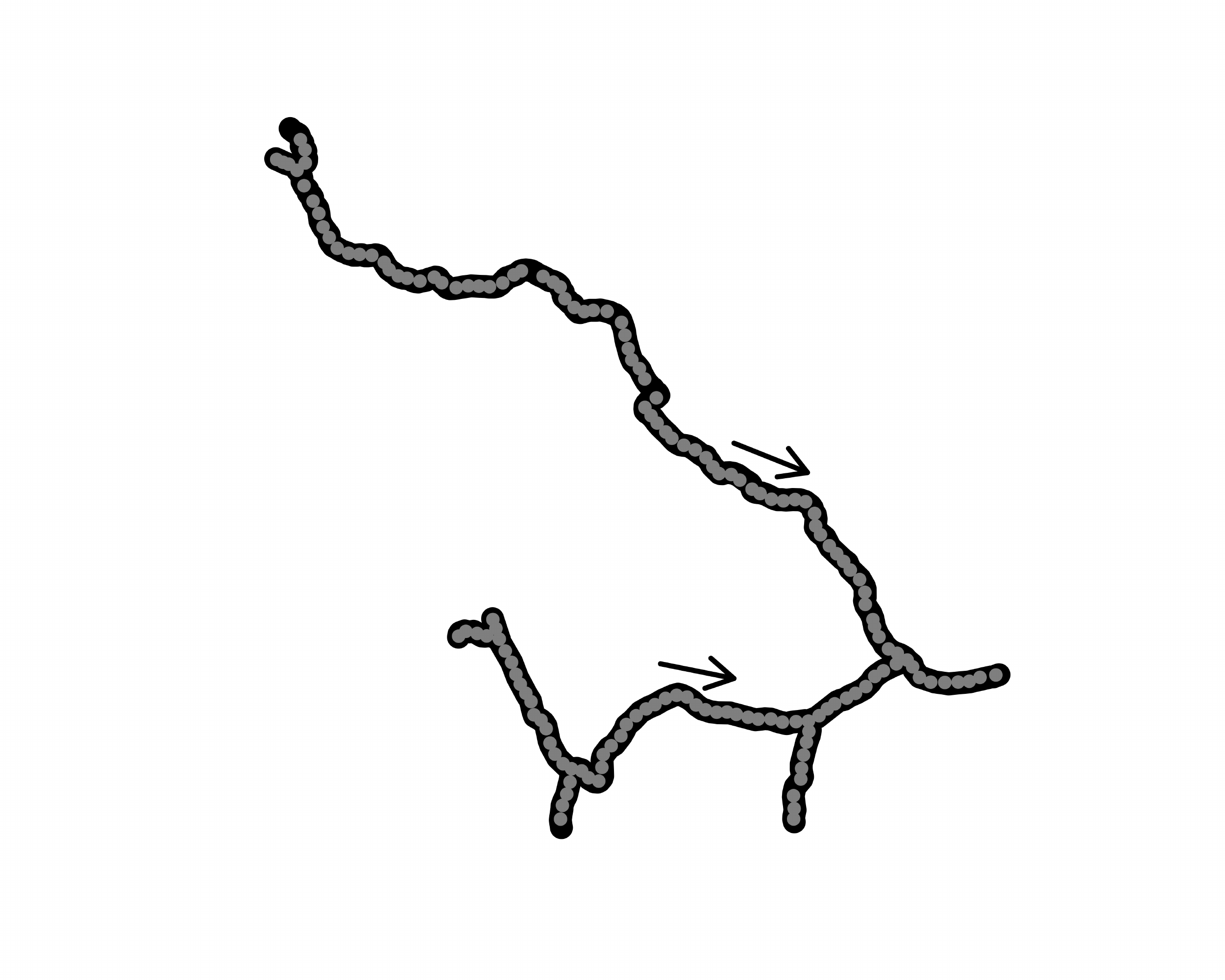}
\end{center}
\caption{Trout sampling locations on the Jefferson Hill Spruce Brook.}
\end{figure}

Additionally, \cite{Kanno2011} examine the effect of two seasonal
blockages of the brook - two locations where the brook dries up and is
seasonally impassible to the trout.  The hypothesized drivers of gene
flow and genetic connectivity among the trout population on Jefferson
Hill Spruce Brook are both directional (differential rates of movement
upstream and downstream) and non-directional (decreased connectivity
between stream locations on opposite sides of the seasonal
blockages).  If spatio-temporal trout movement data were available,
modeling these directional and non-directional responses 
to covariates would be straightforward \citep{Hooten2010b,HanksCTDS}.
For example, movement could be envisioned as occuring on a graph with
a node at each spatial location where trout were 
sampled, and edge weights equal to random walk transition rates
between nodes could be
modeled as
\begin{equation}
\alpha_{ij}=\begin{cases}
\frac{1}{d_{ij}}\text{exp} \left\{\beta_0+\beta_1 u_{ij}+ \beta_2 v_{ij}\right\} & \text{if nodes }i \text{ and }j \text{ are neighbors}\\
0 & \text{otherwise}\end{cases}
\end{equation}
where $\{u_{ij}\}$ and $\{v_{ij}\}$ are indicator variables with
$u_{ij}=1$ if node $j$ is downstream from node $i$ and $v_{ij}=1$ if a
seasonal blockage is located between nodes $i$ and $j$.  In this
formulation, each node on a branch of the stream network has two
neighbors, one upstream and one downstream, and edge weights
$\alpha_{ij}$ are zero for all other non-neighboring nodes.  Each node
at a confluence of two stream branches will have three neighbors, one
downstream and two upstream.  The rate at which a random walker at a
node $i$ on a branch of the stream network transitions to the nearest
upstream node $j$ is $\alpha_{ij}=1/d_{ij}\text{exp}\{\beta_0\}$ if
there is not a seasonal blockage between nodes $i$ and $j$.
Similarly, the rate at which the random walker transitions from $i$ to
the nearest downstream node $k$ is
$\alpha_{ik}=1/d_{ik}\text{exp}\{\beta_0+\beta_1\}$.  The parameter
$\beta_2$ models the additive effect that a seasonal blockage has on
log-transition rates.  Together, this simple random walk model allows
for transition rates that vary with direction and location based on
known spatial stream characteristics.

A spatial model for the observed microsatellite allele
data could then be specified with a latent spatial autocorrelation modeled using the
stationary distribution (11) of the random walk model (12) when driven
by time-homogeneous white Gaussian noise, as described in Section
3.2.  

Microsatellite allele data were observed at $L=15$ distinct loci for
each spatially referenced trout captured.  At the $\ell^{\text{th}}$ locus,
$\ell=1,2,\ldots,L$, denote the
list of all distinct observed alleles from all individuals in the
study as
$\{a_{\ell 1},a_{\ell 2},\ldots,a_{\ell K_\ell}\}$.  Following
\cite{Guillot2005} and others, I model the
two observed alleles for each (diploid) individual as arising from a
multinomial distribution with spatially varying allele probabilities
$\mathbf{p}_{s\ell}=(p_{s\ell 1}\  p_{s\ell 2}\  \ldots \  p_{s\ell
  K_\ell})'$, where $s \in \{1,2,\ldots,S\}$ indexes the spatial
location.

Let $y_{sip\ell k}=1$ if the $p^{\text{th}}$ (indexing ploidy)
observed allele
at the $\ell^{\text{th}}$ locus is
$a_{\ell k}$ for the $i^{\text{th}}$ individual at the $s^{\text{th}}$
spatial location, and $y_{sip\ell k}=0$ otherwise.  Then the
multinomial
probit model \citep[e.g.,][]{Albert1993} for categorical data is often
specified in terms of 
latent variables, $\mathbf{z}$, as follows.
Let
\begin{linenomath}
\begin{equation}
y_{sip\ell k}=\begin{cases}
1 & \ ,\ z_{sip\ell k}=\max\{z_{sip\ell a},\ a=1,\ldots,K_\ell\}\\
0 & \ ,\ \text{otherwise}
\end{cases}
\end{equation}
\end{linenomath}
where 
\begin{linenomath}
\begin{equation}
z_{sip\ell k} \sim N(\mu_{\ell k}+\eta_{s\ell k},1).
\end{equation}
\end{linenomath}

\noindent Then the allele $a_{\ell k}$ makes up a fraction $p_{s\ell
  k}$ of the
genetic makeup of the subpopulation at location $s$, where 
\begin{linenomath}
\[ p_{s\ell k}=\text{Prob}\left(z_{sip\ell k}=\max\{z_{sip\ell a},\
  a=1,\ldots,K_\ell\}\right)\]
\end{linenomath}

\begin{figure}[ht]
\begin{center}
\includegraphics[width=7in,natwidth=576,natheight=216]{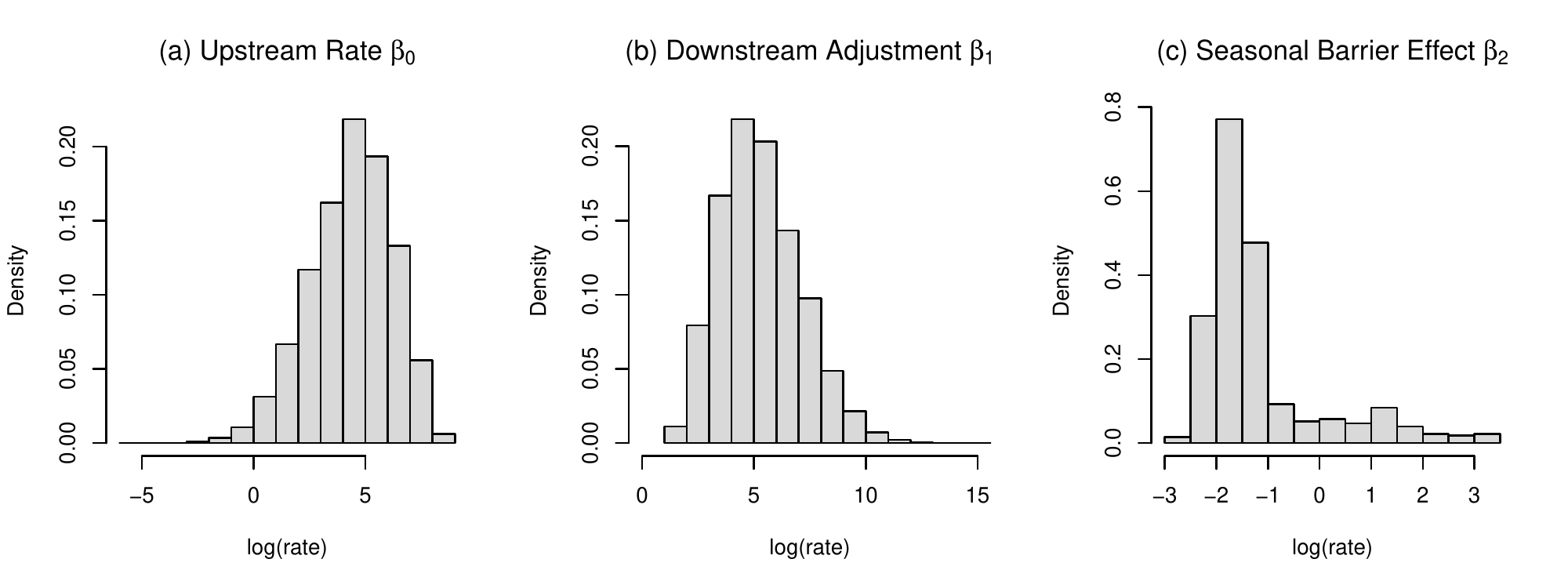}
\end{center}
\caption{Posterior histograms of random walk model parameters in the
  spatial genetic analysis of trout in the JeffersonHill Spruce Brook.}
\end{figure}

The mean of the latent variable $z_{sip\ell k}$ in (14) consists of the
sum of two
effects.  The first is $\mu_{\ell k}$, an allele specific
intercept which determines the relative frequency of the
$k^{\text{th}}$
allele at the $\ell^{\text{th}}$ locus across the entire population
being
studied.  Large values of $\mu_{\ell k}$, relative to $\mu_{\ell k'}$
make it more likely that
$z_{sip\ell k}$ will be larger than $z_{sip\ell k'}$, and so the
$k^{\text{th}}$ allele will be more prevalent than the
$(k')^{\text{th}}$ allele.  Note that the model (13)-(14) is
invariant to a shift in all $\mu_{\ell k}$, as the likelihood is a
function of the  
contrasts  $z_{sip\ell k}-z_{sip\ell k'}$, and not the actual values
of $z_{sip\ell k}$.   Thus, if $\mu_{\ell
  k}$ were replaced by $\mu_{\ell k}+c$ for $k=1,2,\ldots,K_\ell$ and
some constant $c$, the likelihood of the observed allele data would
remain unchanged.  To maintain model identifiability,
fix $\mu_{\ell 1}=0$ for $\ell=1,2,\ldots,L$, as only the relative
differences (contrasts) in $\mu_{\ell k}$ are identifiable.  

The second term in the mean of (14) is $\eta_{s \ell k}$, which is a
spatially varying random effect that allows the allele frequencies
$\mathbf{p}_{s\ell}$ to vary
over the stream network.  Following the reasoning in Section 3.2, the
spatial random effects are modeled as the stationary distribution of a
random walk process driven by time-homogeneous noise.  Let
\begin{linenomath}
\begin{equation}
\boldsymbol\eta_{\ell k}=\left[\begin{array}{c}
\eta_{1\ell k}\\
\eta_{2\ell k}\\
\vdots\\
\eta_{n\ell k}
\end{array}\right]
\sim
N(\mathbf{0},(\mathbf{QQ}')^{-1}),\quad\mathbf{1}'\boldsymbol\eta_{\ell k}=0
\end{equation}\end{linenomath}
where $\mathbf{Q}$ is the infinitessimal generator (9) of the random
walk with transition rates (12).

The model is completed by specifying diffuse Gaussian priors for the
random walk parameters $\beta_0,\ \beta_1,\ \beta_2$ and the allele
specific intercepts
\begin{equation}
\beta_j \sim N(0,10^2),\quad j=0,1,2
\end{equation}
\begin{equation}
\mu_{\ell k} \sim N(0,10^2),\quad \ell=1,2,\ldots,L;\quad k=2,3,\ldots,K_\ell.
\end{equation}

\begin{figure}[htb]
\begin{center}
\includegraphics[width=6in,natwidth=1057,natheight=722]{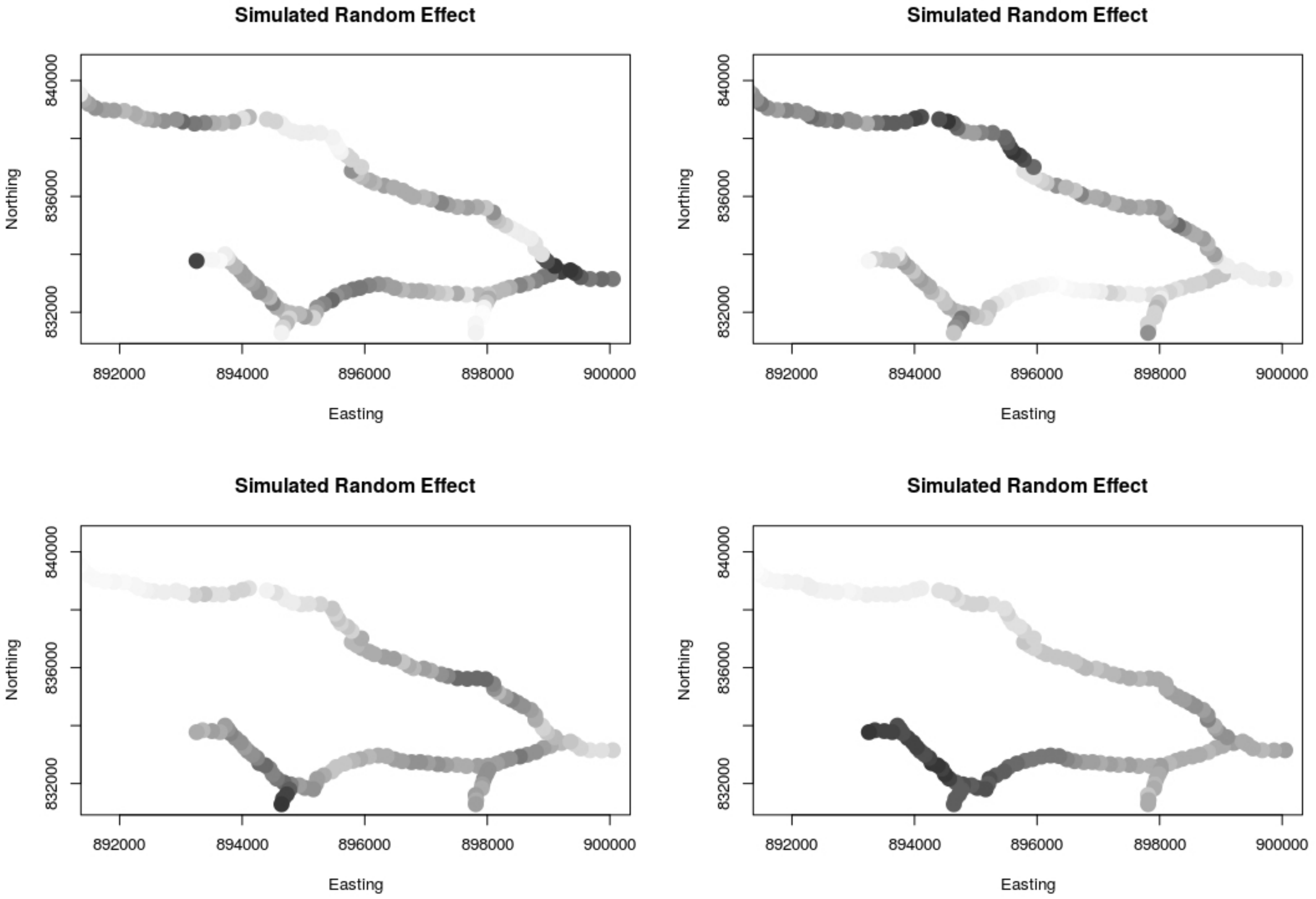}
\end{center}
\caption{Four realizations of random fields on the Jefferson-Hill
  Spruce Brook simulated using posterior mean parameter values of the
  random walk covariance model.}
\end{figure}

A Markov chain Monte-Carlo (MCMC) algorithm was constructed to sample
from the posterior distribution of model parameters, given the
observed microsatellite allele data.  Fifteen chains were run with
different starting values.  Each chain was run for $10^6$ iterations,
with the first $10^5$ samples discarded as burn in.  Convergence was
assessed by comparing posterior histograms obtained from only the first
half of each chain with posterior histograms obtained from only the
second half of each chain.  Histograms of the marginal posterior distributions of the
random walk parameters are given in Figure 3.  The posterior
distribution for $\beta_1$ is greater than zero, indicating that the
data support the anisotropic hypothesis that gene flow is more rapid
downstream than upstream.  The posterior distribution for $\beta_2$,
which captures the effect of the seasonal blockages, overlaps zero
(Figure 3(c)),
with the $95\%$ equal-tailed credible interval being bounded by
$(-2.4,2.2)$.  This indicates only weak support (if any) for the
hypothesis that the seasonal blockages affect gene flow.

Posterior distributions
for the allele specific intercepts are not shown, and posterior mean
values for the intercepts ranged from $-2.2$
to $0.7$.  To qualitatively illustrate the genetic correlation structure
implied by the estimated parameters, four realizations of random
fields on the stream network were simulated using the posterior mean
parameter values.  These random fields are shown in Figure 4.  The
constructive spatio-temporal approach proposed here provides a valid
autoregressive spatial model for data collected on a stream network.
In contrast, \cite{VerHoef2010} present a moving average (convolution)
approach to modeling spatial autocorrelation on stream networks.

\begin{figure}[htb]
\begin{center}
\includegraphics[width=2.61in,natwidth=322,natheight=288]{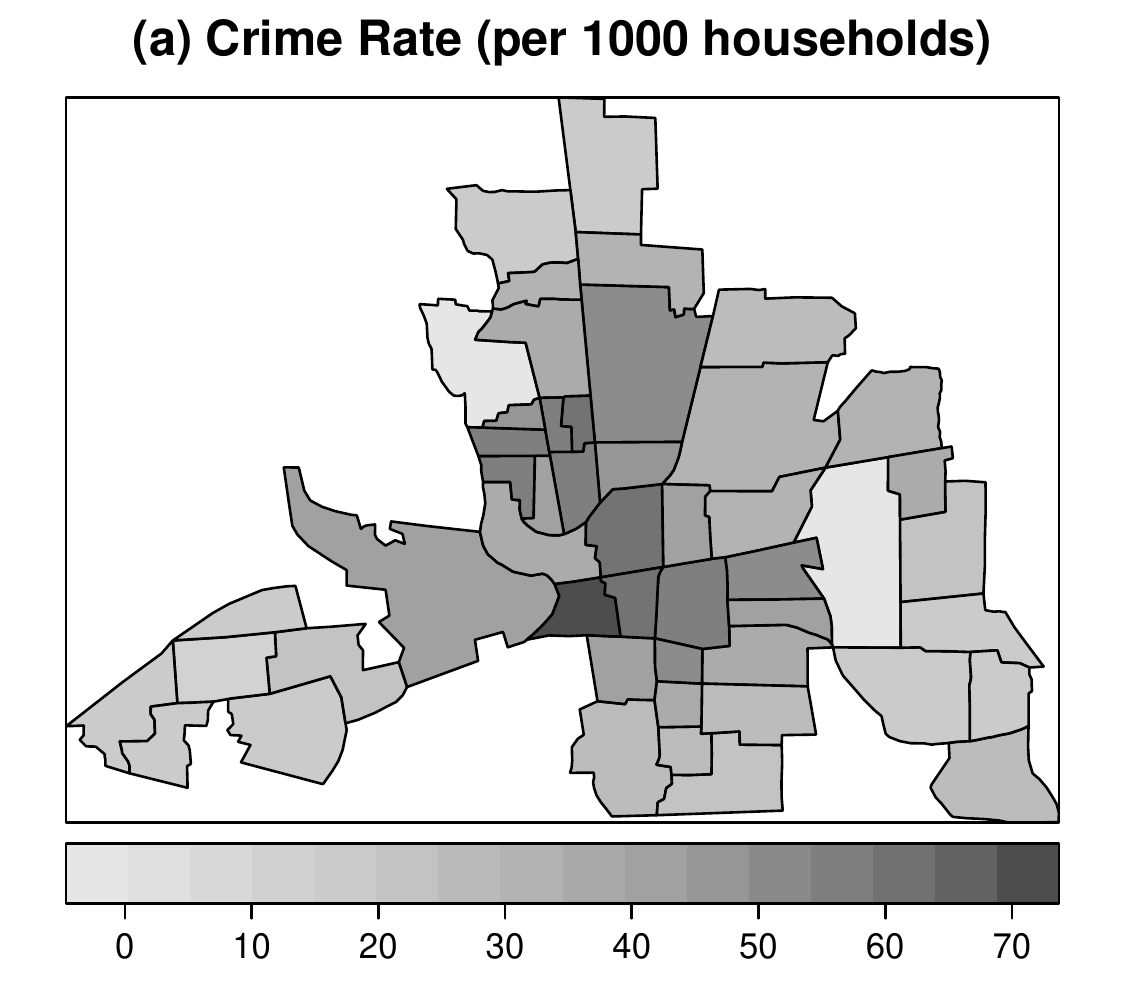}
\includegraphics[width=2.61in,natwidth=322,natheight=288]{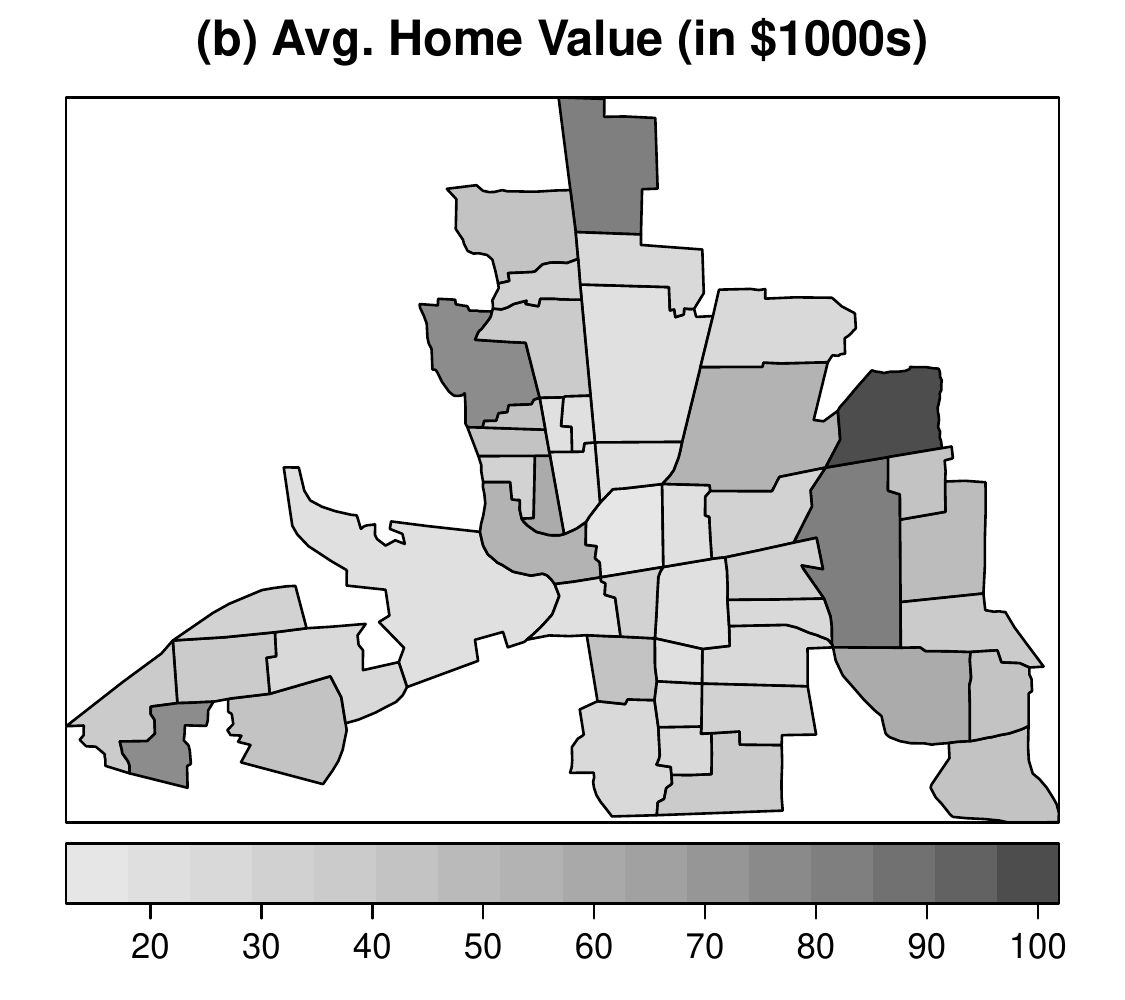}

\includegraphics[width=2.61in,natwidth=322,natheight=288]{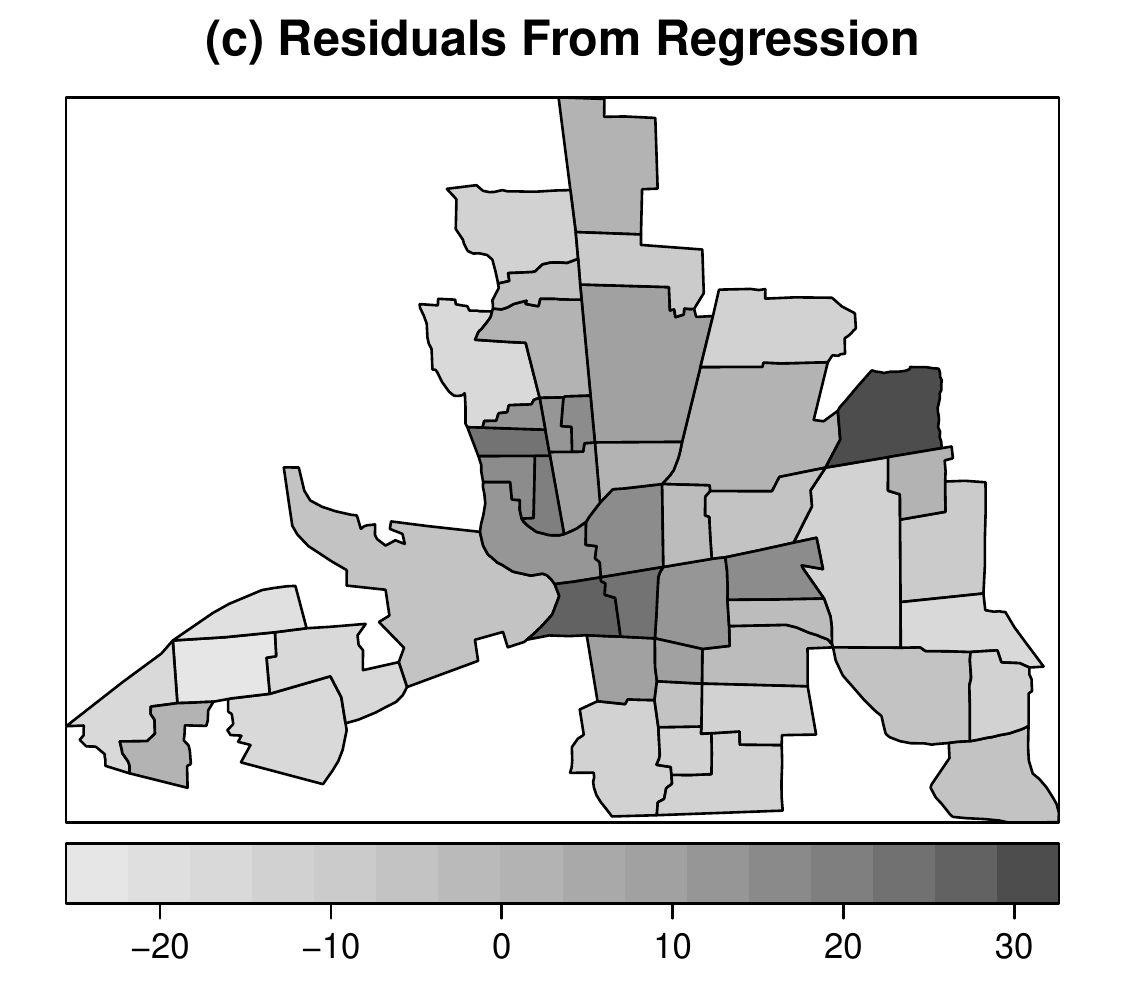}
\includegraphics[width=2.61in,natwidth=322,natheight=288]{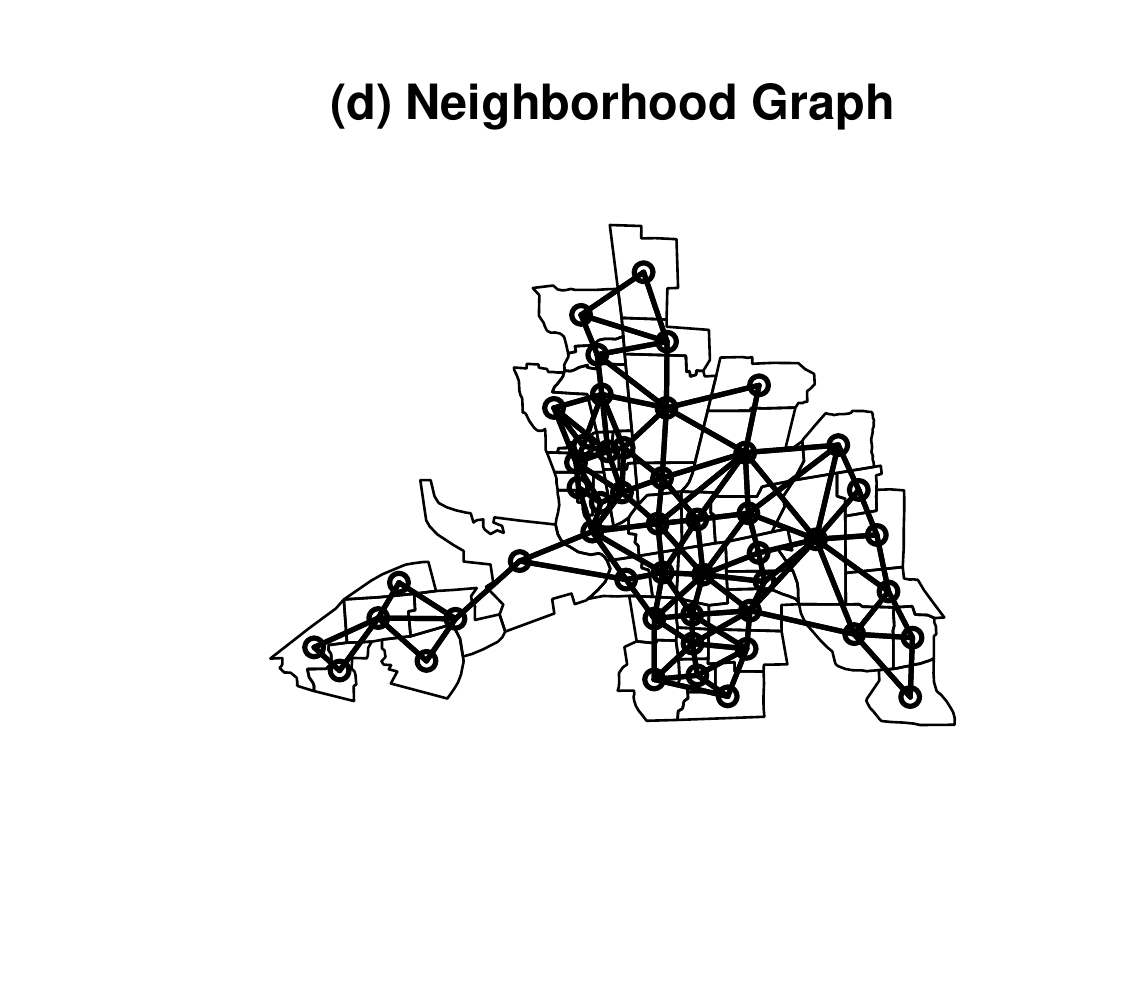}
\end{center}
\caption{Observed 1980 crime rates (a) and average home values (b) in 49 neighborhoods in
  Columbus, Ohio, USA.  The residuals (c) from a simple linear regression
  of crime rates on average home values show clear autocorrelation.  A
standard spatial analysis might include a spatial random effect with SAR
neighborhood structure (d) to account for the spatial
autocorrelation in the data.  We contrast this with a graph diffusion
based approach to jointly modeling spatial autocorrelation and the
effect of the spatial covariate (average home values).}
\end{figure}

\section{Example 2: Crime rates in Columbus, OH.}
 
A second example illustrates how considering a spatio-temporal
generating process can provide insights into modeling the interplay
between mean and covariance structure in spatial models.  As mentioned
previously, recent recognition of the potential for spatial
confounding \citep[e.g., ][]{Hughes2013,HanksRSR} suggests that
correctly modeling the relationship between the fixed and random
effects in a model is important, even if we only desire to interpret
the relationship between fixed effects and the response.

Consider the case of 1980 crime rates in 49 neighborhoods in
Columbus, Ohio, USA \citep{Anselin1988}.  Figure 5(a) shows the number
of residential burglaries and vehicle thefts per thousand housholds
in each of the 49 neighborhoods.  Figure 5(b) shows the average value
of homes in each neighborhood, in thousands of dollars.  These data are freely available in
the `spdep' package \citep{Bivand2015} of the R statistical computing
environment \citep{R}.

A preliminary linear regression with crime rates as response and
average home values as predictor variable indicates a negative
correlation between average home values and crime rates.  However, the
residuals from this simple linear regression are shown in Figure 5(c)
and show clear spatial autocorrelation.  A standard spatial analysis
might consider the following spatial linear model
\begin{equation}
\mathbf{c}=\mu\mathbf{1}+\beta\mathbf{h}+\sigma\boldsymbol\eta+\boldsymbol\epsilon,\quad
\boldsymbol\epsilon\sim N(\mathbf{0},\tau^2\mathbf{I})
\end{equation}
\begin{equation}
\boldsymbol\eta\sim N(\mathbf{0},(\mathbf{QQ}')^{-1})
\text{ with } \mathbf{1}\boldsymbol\eta=0
\end{equation}
where $\mathbf{c}$ is a vector of the 1980 crime rates, $\mathbf{h}$
is a vector of average home values, $\boldsymbol\eta$ is a spatial random effect with SAR structure
defined by $\mathbf{Q}$, and $\boldsymbol\epsilon$ is nonspatial error.  A symmetric neighborhood graph
was defined with edges between all polygons that share a polygon edge, as shown in
Figure 5(d).  If neighborhoods $i$ and $j$ are neighbors, say that
$i\sim j$ or, equivalently in this symmetric relationship, $j\sim
i$. The matrix $\mathbf{Q}$ in (19) then has  
elements 
\[Q_{ij}=\begin{cases}
-1&\ , \text{ if }i\neq j,\ i\sim j\\
0&\ , \text{ if }i\neq j,\ i\nsim j\\
\sum_{j\sim i}1&\ , \text{ if }i=j\end{cases}.
\]
Thus $\boldsymbol\eta$ is an intrinsic spatial random effect with precision
matrix $\mathbf{Q}^2$.  Heuristically, $\boldsymbol\eta$ is a missing
covariate that is spatially smooth on the support of the 49
neighborhoods in Columbus.  

Now contrast this purely spatial approach with an approach based on
considering a spatio-temporal graph diffusion generating process.  As
noted in Section 3.1, the differential equation (8) resulting from
the large $N$ limit of the population-level random walk process is a
 diffusion process defined by a vector of inputs to the system and a
 matrix $-\mathbf{Q}'$ encoding rates of transfer between spatial
 nodes in the graph.  In this spirit, consider a process where the inputs
 (sources and sinks) are random variables with
 mean defined by the predictor variable (average home value) and
 spatial diffusion rates defined by the spatial neighborhood graph.
 Note that while this is not a science-based mechanistic model for
 crime in Columbus, it does provide two competing models for how crime
 rates are related to average home values.  In the standard spatial
 model, the spatial random effect $\boldsymbol\eta$ is a missing
 covariate unrelated to average home values $\mathbf{h}$.  In the
 graph diffusion based model presented below, a diffusion
 process spatially smooths the effect of $\mathbf{h}$, similar to
 a moving average (or convolution-based) spatial model \citep[e.g.,][]{Lee2005}. 

As an alternative to the standard spatial mixed effect model in
(18)-(19), consider modeling crime rates ($\mathbf{c}$) as
\begin{equation}
\mathbf{c}=\mu\mathbf{1}+\boldsymbol\pi+\boldsymbol\epsilon
\end{equation}
where $\boldsymbol\pi$ is the stationary distribution of the
spatio-temporal graph diffusion process $\mathbf{z}(t)$ defined elementwise as
\begin{equation}
\frac{\partial z_i(t)}{\partial t} = -\kappa n_i z_i(t)
+\sum_{j\sim i} \kappa z_j(t)  +
  \beta\cdot h_i+\delta_i.
\end{equation}
The first term on the right hand side of (21) defines the flow out of
node $i$ to the $n_i=\sum_{j\sim i} 1$ neighboring nodes.  The second term defines the flow into node
$i$ from other nodes.  The net input/output from ``births'' and
``deaths'' into node $i$ is $\beta
h_i+\delta_i$.  The intuition here is that the spatial source of crime in
Columbus neighborhoods is correlated with home values, and that crime
spreads spatially out from neighborhoods with high crime rates to
neighboring regions, with a constant diffusion rate of $\kappa$
between all neighboring nodes.

If $\delta_i$ are modeled as independent zero mean Gaussian
random variables, the RPDE can be written in vector form as 
\begin{equation}
\frac{\partial \mathbf{z}(t)}{\partial t} =
-\kappa\mathbf{Q}'\mathbf{z}(t)+ \beta\mathbf{h}+ \boldsymbol\delta,\quad
\boldsymbol\delta\sim N(\mathbf{0},\sigma^2\mathbf{I})
\end{equation}
and the stationary distribution $\boldsymbol\pi$ satisfies
\[\kappa\mathbf{Q}'\boldsymbol\pi=\beta\mathbf{h}+\boldsymbol\delta,\]
or, equivalently
\[\boldsymbol\pi \sim
N\left(\frac{\beta}{\kappa}(\mathbf{Q}')^{-1}\mathbf{h},\frac{\sigma^2}{\kappa}(\mathbf{QQ}')^{-1}\right),\quad
\mathbf{1}'\boldsymbol\pi=0\]
where $(\mathbf{Q}')^{-1}$ is the Bott-Duffin constrained generalized
inverse \citep{Bott1953} of $\mathbf{Q}'$.
The data model (20) for the graph diffusion spatial model can then be
written as
\begin{equation}
\mathbf{c}=\mu\mathbf{1}+\tilde{\beta}(\mathbf{Q}')^{-1}\mathbf{h}+\tilde{\sigma}\boldsymbol\eta+\boldsymbol\epsilon
\end{equation}
with $\tilde{\beta}=\beta/\kappa$, $\tilde{\sigma}=\sigma/\kappa$, and
$\boldsymbol\eta$ a random effect defined as in (19).  Without strong prior
information, $\kappa$ will be unidentifiable.  Instead, consider
inference on $\tilde{\beta}=\beta/\kappa$ and
$\tilde{\sigma}=\sigma/\kappa$, which are identifiable.  In this
formulation, the only difference
between the standard spatial model in (18) and the graph diffusion
based spatial model in (23) is that the fixed effect $\mathbf{h}$ in
(18) is smoothed by $(\mathbf{Q}')^{-1}$ in (23).      

Within a Bayesian framework for inference, I assigned flat Gaussian
priors to the regression parameters $\mu$, $\beta$, and
$\tilde{\beta}$.  Flat half-normal priors were chosen for the spatial random effect variance parameters
$\sigma$ and $\tilde{\sigma}$, and an inverse-gamma prior was chosen
for the non-spatial error variance $\tau^2$.  Inference on the parameters in (19) and (23) was obtained by a
Markov chain Monte Carlo sampler.  In each case, the MCMC sampler was
run for $10^5$ iterations.  Convergence was assessed by comparing
histograms of samples from the first half of the Markov chain with
histograms of samples from the second half of the Markov chain.  

\begin{table}[htb]
 \centering
\caption{Posterior results for parameters in the spatial and
  graph-diffusion based models for crime in Columbus, OH
  neighborhoods.  The graph diffusion model fits the data better as
  measured by DIC.}
 \begin{tabular}{rrrr}
   \hline
 Parameter & Post. Mean & Post. 0.025 Quantile & Post. 0.975 Quantile \\
   \hline
\multicolumn{4}{l}{\textbf{Spatial Model (18) $\quad$ DIC = 442.10}}\\
\hline
 $\mu$  & 35.12 & 32.09 & 38.12 \\
 $\beta$ & -9.28 & -12.48 & -6.16 \\
 $\sigma$ & 1.81 & 0.31 & 3.50 \\
 $\tau$ & 10.75 & 8.86 & 13.04 \\
   \hline
\multicolumn{3}{l}{\textbf{Graph Diffusion Model (23) $\quad$  DIC = 411.52}}\\
\hline
 $\mu$ & 35.13 & 31.89 & 38.33 \\
 $\tilde{\beta}$ & -9.38 & -12.89 & -5.92 \\
 $\tilde{\sigma}$  & 0.94 & 0.03 & 2.67 \\
 $\tau$ & 11.51 & 9.68 & 13.75 \\
    \hline
 \end{tabular}
 \end{table}
  
Posterior means and 95$\%$ credible interval bounds are shown in Table
2.  To compare models, I computed the Deviance information criterion
\citep[DIC, ][]{Spiegelhalter2002}.  Posterior distributions for $\mu$
and $\beta$ from the spatial model (18) are similar to those of $\mu$
and $\tilde{\beta}$ from the graph diffusion model (23); however, the
standard deviation $\sigma$ of the spatial random effect
$\boldsymbol\eta$ in the spatial model (18) is larger than the
corresponding standard deviation $\tilde{\sigma}$ in the graph
diffusion model (23).  This indicates that the need for the spatial
random effect is greater in the spatial model than in the graph
diffusion model where the home value covariate was smoothed by
$(\textbf{Q}')^{-1}$.  The DIC of the graph diffusion model (DIC=411)
was lower than that of the standard spatial model (DIC=442), indicating that in
this case, considering a spatio-temporal generating process resulted
in a better model fit than would be obtained by the inclusion of a
standard spatial random effect.

\section{Discussion}

While we have focused on discrete space models, this general approach
has potential for application in continuous space as well.  Spatial deformation
approaches to nonstationary covariance \citep[e.g.,
][]{Schmidt2003,Lindgren2011} can be viewed as stationary
distributions of diffusion processes with spatially heterogeneous
diffusion rates.  Reaction-diffusion models are common in ecology and
other fields \citep[e.g.,][]{Keeling2004, Hu2013} and would provide a natural
spatio-temporal generating process basis for spatial random effect
models in a wide variety of systems.  Finite element basis and
grid-based 
approaches to approximating continuous spatial fields have a long
history in spatio-temporal \citep[e.g.,][]{Wikle2010} and spatial
\citep[e.g.,][]{Lindgren2011} analysis, and could be used to
approximate the stationary distribution of a continuous
(infinite-dimensional) spatio-temporal
generating process with a finite number of basis functions.

Current standard approaches to modeling spatial correlation focus on
nonparametric random effect models.  This work proposes a
parametric constructive approach to modeling spatial random effects based on
an assumed spatio-temporal generating process.  The two examples give
some indication of how this approach may be used.  In the first
example, existing scientific knowledge about the system (gene flow on
a stream network) was used to specify a spatio-temporal generating
model (a population-level random walk), and the stationary
distribution of this spatio-temporal process defined the distribution
of the spatial random effect used to model genetic correlation.  In
the second example, a descriptive approach was taken to compare
multiple models for spatial variation.  In particular, for 
the Columbus crime data, the graph diffusion model provided a better
model fit than was obtained using a standard spatial random effect
model.  Modeling spatial random effects nonparametrically is the
current standard practice; however,
there are benefits to parametric modeling of spatial random effects
when the existing science can suggest a spatio-temporal generating
mechanism.

\appendix

\section*{Appendix A: Large population limits of population processes}

The interested reader is referred to \cite{Kurtz1981} for a full
treatment of stochastic population processes.  This derivation follows
the spirit of \cite{Kurtz1981} and \cite{Baxendale2011}, but with the
novelty of birth and death rates that are not density dependent.

Following from (6) in Section 3.1, the transient distribution for the normalized density
$\mathbf{z}=\mathbf{n}/N$ is given by
\begin{equation*}
\mathbf{z}(t) = \mathbf{z}(0)+ \sum_{ij \neq 0}
(\mathbf{e}_{j}-\mathbf{e}_i) \frac{1}{N}
P_{ij}\left[ \int_0^t n_i(s)\alpha_{ij}\text{d}s\right] 
+ \sum_i \mathbf{e}_i   \left(\frac{1}{N}P_{0i}\left[ Nb_i t\right] - \frac{1}{N}P_{i0}\left[ Nd_i t\right]\right)
\end{equation*}
where \[P_{ij}(a) \sim Pois(a),\quad i=0,1,\ldots,M;\ 
j=0,1,\ldots,M; \ i\neq j.\]

Note that 
\begin{align*}
P_{ij}(a) &= a +(P_{ij}(a)-a)\\
&=a+W_{ij}(a)\ ,\quad \quad W_{ij}(a)\sim(0,a)
\end{align*}
where each $W_{ij}$ has mean zero on constant variance.  Applying this
to the transient distribution gives
\begin{align*}
\mathbf{z}(t) &= \mathbf{z}(0)+ \sum_{ij \neq 0}
(\mathbf{e}_{j}-\mathbf{e}_i) \frac{1}{N}
\left[ \int_0^t n_i(s)\alpha_{ij}\text{d}s\right] 
+ \sum_i \mathbf{e}_i   \left(b_i t-d_i t\right) \\
&+\frac{1}{N}\left(\sum_{i \neq j} (\mathbf{e}_{j}-\mathbf{e}_i)
W_{ij}\left[ \int_0^t n_i(s)\alpha_{ij}\text{d}s\right] + \sum_i
\mathbf{e}_{i} \left(W_{0i}\left[ Nb_i t\right] - W_{i0}\left[ Nd_i t\right]\right)\right).
\end{align*}

Consider a fixed $t>0$ and note that $N \geq n_i(s)$ for all $s\in
(0,t)$.  This gives the result that 
\[\int_0^t n_i(s)\alpha_{ij}\text{d}s \leq N\alpha_{ij}t.\]
Then to show that all terms above including random variables
$W_{ij}$ disappear in the
limit as $N\rightarrow \infty$, it is enough to consider the behavior
of 
\[\frac{1}{N}W(Na),\quad W(a)\sim (0,a)\]
for a constant $a>0$.  It is trivial to note that 
\[E\left[\frac{1}{N}W(Na)\right]=0\]
and that 
\[Var\left[\frac{1}{N}W(Na)\right]=\frac{1}{N^2}Na\]
which vanishes in the limit as $N\rightarrow \infty$.

Then, in the large population limit, the transient distribution of the
normalized population $\mathbf{z}(t)$ will be given by 
\[\mathbf{z}(t) = \mathbf{z}(0)+ \sum_{i \neq j}
(\mathbf{e}_{j}-\mathbf{e}_i) \frac{1}{N}
\left[ \int_0^t n_i(s)\alpha_{ij}\text{d}s\right] 
+ \sum_i \mathbf{e}_i   \left(b_i t-d_i t\right). \]

\section*{Appendix B: Proof of Theorem 3.1}

In this appendix, we prove Theorem 3.1.  
The proof follows from the fact that $\mathbf{QQ}'$ is a
Gramian matrix \citep[e.g.,][]{GentleText} and thus
$\mathbf{QQ}'=\mathbf{WW}'$ if and only if
$\mathbf{W}=\mathbf{QU}'$ for a real unitary matrix
$\mathbf{U}'$.  As $\mathbf{W}$ and $\mathbf{Q}$ are both generators
for CTMC random walks, their rows sum to zero
($\mathbf{Q1}=\mathbf{W1}=\mathbf{0}$), with negative diagonal entries
($q_{ii}<0$, $w_{ii}<0$) and non-negative off-diagonal entries
($q_{ij}\geq0$, $w_{ij}\geq 0$ for $i\neq j$).  If $\mathbf{Q}$ and
$\mathbf{W}$ are both generators for irredicible CTMCs, then both
matrices have rank
$n-1$ and their null spaces are both spanned by the $\mathbf{1}$
vector.  As $\mathbf{W1}=\mathbf{0}$, it follows that
$\mathbf{QU}'\mathbf{1}=\mathbf{0}$ and thus
$\mathbf{U}'\mathbf{1}=\lambda\mathbf{1}$ for some $\lambda$.  The
eigenvalues of any unitary matrix $\mathbf{U}'$ have absolute value
equal to 1, so $\lambda$ either equals $1$ or
$-1$.  If $\mathbf{u}'_i$ is the $i$-th row of $\mathbf{U}'$, then
$\mathbf{u}'_i\mathbf{1}$ equals either 1 or $-1$, but since
$\mathbf{U}$ is unitary, $\mathbf{u}'_i\mathbf{u}_i=1$.  These requirements both hold if and only
if  $\mathbf{u}_i=\lambda \mathbf{e}_k$, where $\mathbf{e}_k$ is the
canonical vector with $k$-th element equal to 1 and all other elements
equal to zero.  As $\mathbf{U}$ is of full rank, the rows of
$\mathbf{U}'$ must contain a full set of canonical vectors spanning
$\mathcal{R}^n$. 

First consider the case where $\lambda=1$.  Then $\mathbf{U}'$ is a
permutation matrix, with the columns of $\mathbf{W}$ being permuted
columns of $\mathbf{Q}$.  However, as $\mathbf{W}$ and $\mathbf{Q}$ are generator
matrices, each diagonal entry of $\mathbf{W}$ and $\mathbf{Q}$ must be
negative, while 
all off-diagonal entries are non-negative.  This can only hold for
$\mathbf{W}$ if the
permutation matrix $\mathbf{U}'$ is the identity matrix, and thus
$\mathbf{W}=\mathbf{Q}$.  

Now consider the case where $\lambda=-1$.  Again $\mathbf{U}'$
permutes the columns of $\mathbf{Q}$, but now the sign of all entries
is changed through multiplication by $\lambda=-1$.  So
$w_{ii}=-q_{ik}$ and $w_{ik}=-q_{ii}$ for some $k$.  As $\mathbf{W}$
is a generator matrix, $w_{ii}=-\sum_{j\neq i} w_{ij}$, which is only
possible if $q_{ik}$ is the only non-zero off-diagonal entry in the
$i$-th row of $\mathbf{Q}$.  This completes the proof.


\bibliographystyle{agsm}

\bibliography{/home/ephraim/Dropbox/Research/library}

\end{document}